\documentclass{svproc}

\usepackage{url,booktabs}

\usepackage{graphicx}
\usepackage{hyperref}
\hypersetup{
    colorlinks=true,
    linkcolor=blue, 
    urlcolor=blue,
}

\begin{document}
\mainmatter              % start of a contribution

\title{Ethics and EU AI Act in Cases of Work Disability Risk and Alzheimer's Disease Risk Prediction}
\titlerunning{Ethics and EU AI Act in Cases}  % abbreviated title (for running head)

\author{*Sami Andberg\inst{1} \and Henri Terho\inst{2} \and Katja Saarela\inst{2}}
\authorrunning{Andberg, Terho, and Saarela} % abbreviated author list (for running head)

\institute{University of Eastern Finland, Joensuu, Finland,\\
\email{saman@student.uef.fi},\\
\and
Eficode Group Ltd, Helsinki, Finland,\\
\email{henri.terho@eficode.com},\\
\email{katja.saarela@eficode.com}}

\maketitle              % typeset the title of the contribution

\begin{abstract}

Improvements in AI technologies have made it feasible to develop new types of medical AI tools. However, these tools raise new kinds of questions, especially in relation to the ethics and AI Act compliance. We analyzed two cases of AI tools developed to predict medical risks, the risk of work disability (case A) and the risk of getting Alzheimer's disease (case B). We observed both cases using the ethical AI and the EU AI Act as frameworks, noted that they classify as high-risk systems, and that bringing them from the research environment to production would require a lot of work and compliance due to the related regulation. \par


\end{abstract}
\textbf{Acknowledgement:} This preprint has not undergone peer review or any post-submission improvements or corrections. The Version of Record of this contribution is published in Lecture Notes in Networks and Systems, volume 1907, The 2026 International Conference on Breakthroughs in Artificial Intelligence (BAI’26), and is available online at \url{https://doi.org/10.1007/978-3-032-21147-7\_19}.

\section{Introduction}
Artificial Intelligence (AI) has been one of the big breakthroughs in the recent years. Almost every week we hear about a new product, new model, or a new capability AI can do, and companies are rushing to integrate AI into their operations while students and regular users have picked it up as a part of their daily routines. The speed of the adaptation of the new technology has caught most by surprise, and sparked debate on the ethics and regulations related to the safe and beneficial use of AI. Ethics is a core topic in both Medical research and AI. Many of the topics relating to both these fields are impactful topics requiring careful study of their ethical impacts to avoid maleficence or misuse. Systematic literature review by Kargl et al. shows the need for an ethical, mindful and balanced approach to AI in biomedical research, specifically the need for AI ethics research \cite{kargl2022literature}.

One way to handle the ethical aspects is to use the AI for Good framework. AI for the Good refers to methods through which AI is used for beneficial purposes \cite{vieweg2021ai}. An AI for the Good framework is AI4People, which was established to help use AI towards the good of society, its members, and the environment \cite{AI4}. An ideal society could be one where as many people as possible remain healthy and capable of working. The tax money collected from work would take care of children, the elderly, and the people who are unable to work themselves. However, we do not live in an ideal society: Too many people become unable to work before their actual retirement age \cite{OECD}. Especially, various types of dementia, such as Alzheimer’s disease, have increased significantly \cite{zeisel2020world}. It is also known that the earlier a person receives support, the more likely they are to avoid disability pensions and to be able to continue in their job. The same applies to dementia: early intervention slows the progression of the disease and provides additional healthy years \cite{zeisel2020world}. The early detection of various diseases is therefore important.

At the same time, Artificial Intelligence (AI) methods have developed tremendously. We can describe AI as systems that display intelligent behaviour by analyzing their environment and taking actions -- with some degree of autonomy -- to achieve specific goals \cite{sheikh}. Different AI techniques are, for example, rule-based systems and machine learning (ML) \cite{hoffman}. AI enables large amounts of data to be processed efficiently. It is possible to develop methods that predict disease risk and facilitate early intervention. Thus, we have developed AI methods for predicting work disability risk \cite{huhta2020,saarela2022,huhta2023,saarela2023ethical,saarela2023explainability,saarela2023book,saarela2025} and for the early detection of Alzheimer's disease \cite{andberg2025pupil,hannonen2022shortening,shah2025multimoda}. There are still legal and ethical aspects that we have to take into account. In Europe we need to follow EU AI Act \cite{EU}, which categorises AI systems into different categories based on their risk level. The following Research Questions guides this study:\\

RQ1: What are the risk categories of our AI Systems for work disability and Alzheimer's disease risk prediction from EU AI Act and ethical risks?\\

RQ2: How do these risk categorizations affect our AI system development and deployment?\\

In this article, we will first briefly discuss AI Ethics and EU AI Act. Secondly, we will go through the two AI system cases and investigate what are their risk categories in EU AI Act \cite{EU}. Finally, we will discuss the results and give conclusions. 

\section{AI Ethics and EU AI Act}

Major collection of principles for ethical medical research is the Declaration of Helsinki \cite{helsinki2024}. It works as the global standard for medical research targeting humans, focusing on four core medical ethics: beneficence, non-maleficence, respect for autonomy and justice. These principles are supported by the Belmont report \cite{belmont1979} from the US and codified into guidance in the CIOMS Guidelines \cite{cioms2016}. These principles form a system, which is patient right centric in medical research. 

The second column of ethics in medical AI research is formed by ethics around AI research. 
Jobin et al. \cite{jobin2019} have formulated a model of five aspects of ethical AI: non-maleficence, accountability and responsibility, transparency and explainability, justice and fairness, and respect for various human rights. These aspects should be considered when collecting data, storing it in databases, and sharing it with others. 

Same aspects are also in EU AI Act \cite{EU}. It’s the first comprehensive law in the world that specifically targets AI, and it aims to balance innovation, safety, and fundamental rights. The aim is to promote trustworthy AI while still encouraging innovation. China \cite{china} and the USA \cite{usa} have also launched similar types of governance initiatives. These aim to legally control and burden research to ensure compliance and lawfulness in research, bringing in, in the case of AI Act, the EU AI Office as the highest level supervisor to monitor compliance. This shifts the style of ethical control from distributed Independent review boards (IRBs) to centralized EU Office bureacracy. 

EU AI Act \cite{EU} is a risk based approach that divides AI systems into four categories: Unacceptable Risk, High Risk, Limited Risk, and Minimal Risk. AI systems with unacceptable risks are, for example, social scoring by governments, AI manipulating vulnerable groups, and real-time biometric surveillance in public. Systems with significant impact on health, safety, or rights belong to high risk category. Examples of high risk AI systems are AI in medical devices, hiring and recruitment algorithms, credit scoring and critical infrastructure. Chatbots and generative AI tools are examples of limited risk AI systems whereas spam filters and using AI in video games are examples of minimal or no risk AI systems. Thus the way EU AI Act handles ethics is more focused on the system, not the user. Or in our case, patient rights.

The interaction of the EU AI Act with medical research thus changes the focus of medical research ethics from patient centric to system centric through it, by codifying ethics principles into EU ethics law. Previously medical ethics allowed flexibility for innovation, if consent and oversight are respected, but AI act controls this through risk categorization.

These three columns are outlined in Table \ref{table-ethics}. We collect and compare the different features of the different ethics to their intreprepration in the AI Act.

\begin{table}[ht!]
\centering
\caption{Comparison of Medical Research Ethics, AI Ethics Guidelines, and EU AI Act}
\label{table-ethics}
\scalebox{0.79}{
\begin{tabular}{p{3cm} p{4cm} p{4cm} p{4cm}}
\toprule
\textbf{Principle / Focus} & \textbf{Medical Research Ethics} & \textbf{AI Ethics Guidelines} & \textbf{EU AI Act} \\
\midrule
\textbf{Base Sources} & 
Declaration of Helsinki, Belmont Report, CIOMS Guidelines & 
EU HLEG Guidelines on Trustworthy AI, OECD AI Principles, Jobin et al. (2019) & 
Regulation (EU) 2024/1689 (AI Act) \\
\midrule
\textbf{Core Ethical Principles} & 
Respect for persons, beneficence, justice, informed consent & 
Human agency, non-maleficence, fairness, transparency, accountability, human rights & 
Trustworthy AI via legal compliance: safety, transparency, accountability, fairness \\
\midrule
\textbf{Primary Focus} & 
Protection of individual research participants & 
Trustworthy design and deployment of AI systems & 
Risk-based regulation of AI systems across society \\
\midrule
\textbf{Risk–Benefit Balance} & 
Scientific validity required, risk minimized, benefits must outweigh harm & 
Technical robustness, avoidance of bias, non-maleficence & 
High-risk AI requires assessment, documentation, and human oversight \\
\midrule
\textbf{Oversight Mechanism} & 
Ethics committees, independent review boards (IRBs) & 
Voluntary principles and guidelines, self-regulation & 
Binding legal requirements, national authorities, EU AI Office supervision \\

\midrule
\textbf{Flexibility vs. Compliance} & 
Ethics allows proportional risk if oversight and consent are secured & 
Non-binding, aspirational frameworks; innovation encouraged & 
Strict compliance obligations for high-risk AI (bias audits, monitoring, documentation) \\
\midrule
\textbf{Impact on Research} & 
Facilitates responsible innovation through consent + IRB review & 
Encourages transparency and accountability without legal enforcement & 
May slow translation of health AI from research to clinical practice due to regulatory burden \\
\bottomrule
\end{tabular}
} %close scalebox
\end{table}

\section{Case Studies}
In this section, we will present two cases of developing AI based systems in a medical setting, and focus on the aspects presented in previous section, namely medical ethics and AI ethics. Case A concerns work disability risk prediction and Case B handles Alzheimer's disease prediction. 

\subsection{Case A: Work disability risk prediction}
We developed a machine learning method $M_{Health}$ for work disability risk prediction \cite{huhta2020,saarela2022,huhta2023,saarela2025}. It is a Natural Language Processing (NLP) method based on medical records.  Architecture of the language model is based on the ASGD Weight-Dropped Long Short Term Memory network (AWD-LSTM) that is a variation of recurrent neural network (RNN) \cite{merity2017}. A corpus of anonymized medical reports is collected from patients' visits to an occupational health doctor. There are two classification groups: no disability risk and disability risk. Medical experts labeled texts to the risk and non-risk classes and the algorithm tried to learn the classification. We also developed a three-class model where classes are no risk, risk, and high risk. The model is created based on FastAI's ULMFiT \cite{howard2018} using 6000 labeled Finnish medical records as training data and 3000 records for validation. 

We have studied ethical aspects of the model \cite{saarela2023ethical,saarela2023explainability}. These sources cover the columns of both medical and AI ethics in Table \ref{table-ethics}. Where we account for the risk-benefit balance, oversight mechanism and asses the core ethical principles outlined by Jobin. This shows that the oversight mechanisms of independent review have been used in the study \cite{jobin2019}. 

But these earlier papers have not studied how our method is classified in EU AI Act \cite{EU} and how the legislation affects to developing and using that kind of systems. Because our AI predicts work disability risk in the healthcare sector, it involves categorisation of people. EU AI Act \cite{EU} refers several times to biometric categorisation and emotion recognition but not mention directly our Case A where we predict the risk using textual data. However, EU AI Act defines characteristic of "AI system" (definition 12) their capability to infer using the process of obtaining the outputs, such as predictions, content, recommendations, or decisions. The output in Case A a is prediction. 

\subsection{Case B: Alzheimer's disease risk prediction}
The initial BEGAD study collected multimodal and diagnostic data for 78 elderly participants in Eastern Finland, 21 of whom had mild Alzheimer's disease (AD) diagnosis while 57 were non-demented. After cognitive tests and medical expert board review 20 of the non-demented group were re-classified as having Mild Cognitive Impairment (MCI). The data collected included cognitive tests (CERAD-NB, MMSE) as well as an eye-tracking study where participants the read numbers and text displayed on the screen aloud as fast as possible \cite{hannonen2022shortening}.

We analysed the BEGAD-2019 eye-tracking dataset using different methods - statistical, machine learning, and deep learning - to find the best techniques and features to detect significant differences between groups with the goal of being able to develop an AI tool to identify the individuals at the highest risk of developing dementia. Statistical analysis based of the traditional eye-tracking features - fixations and saccades - found group-wise differences in participants' saccade amplitudes and durations \cite{hannonen2022shortening}, while the machine learning approach (using a larger feature set) emphasized baselined pupil size -derived features over others as most promising candidates for classifying the participants \cite{andberg2025pupil}. The deep learning study utilized an even larger multimodal dataset also including participants' transcribed audio data which was processed along the eye-tracking data, thus allowing multimodal fusion features. We evaluated the classification power of the features in this multimodal feature set using various machine learning and deep learning tools. The pupil size -derived features, as well as multimodal features fusing speech and gaze data, were the main contributors to the classification power in the best-performing FNN-model \cite{shah2025multimoda}. 

While the BEGAD study data collection was processed and approved by the relevant hospital district, the specific AI tools developed for the BEGAD data analysis haven't been evaluated in this aspect before. Referring to the two first columns - Medical research ethics and AI ethic goudelines - of Table \ref{table-ethics}, we can conclude that our system seems to balance both aspects. In relation to the participants, the focus is on the persons, they gave informed consent to participate in the data collection (or a proxy consent was given for the participants already diagnosed with AD), and the goal is to benefit the future users. On the AI side, the goal was to develop a system that works truthfully and fairly, shows transparent results (in a way that it lists which features helped it reach the conclusion) and works as an assisting technology to the clinician making the final diagnosis. The aforementioned points match the first and second column respectively.

\section{Results and Discussion}

Assessing the ethical aspects of an AI system is not always straightforward. One has to take into account the multiple aspects and reference groups related to the system; developers, administrators, specialists, doctors, the subjects themselves, as well as regulators and the associated technical providers. It is important to note that in both of these systems, AI has a supportive role. The expert is responsible for the decisions, not AI. 

In this work we evaluated the two cases we were familiar with in relation to these aspects. Principles for ethical medical research \cite{helsinki2024} bring patient to the center, whereas EU AI Act is concentrating on risk- and system-based approach. We suggest that the more relaxed medical research ethic approach would be beneficial for our cases in bringing them out for use as compared to the more heavily regulated EU AI Act's viewpoint.

%We also set out to examine how our systems would be considered 
 As both work disability risk prediction and Alzheimer's disease risk prediction categorize people, they belong to the High-Risk category of EU AI Act (Definition 96) \cite{EU}. The result was the same for both Provider and Developer roles. The summary of the methods can be found in Table \ref{table-features}.

\begin{table} [ht!]
    \caption{\label{table-features} Comparison of the Cases}
      \begin{tabular}{lll}        
        \toprule
        &Case A&Case B\\
        \midrule
        Purpose&Work disability risk prediction&Alzheimer's disease risk prediction\\
        Method&Deep learning/NLP&Deep Learning/FNN\\
        Data&Medical records&Biometric, diagnostic and audiovisual data\\
        Risk Category&High Risk&High Risk\\
        \bottomrule
    \end{tabular}
\end{table}

According EU AI Act \cite{EU}, as a provider of a high-risk healthcare AI system, one must carry out a conformity assessment before market placement. One is  also responsible for bias and fairness testing, documentation of data and design choices, and human oversight in deployment. In addition to that, post-market monitoring is required to track system performance and risks over time. In this regard, the EU AI Act (Definition 96) states that "The European Artificial Intelligence Office (AI Office) should develop a template for a questionnaire in order to facilitate compliance and reduce the administrative burden for deployers." 

According to the EU AI Act (Definition 25 \cite{EU}), this regulation should support innovation, respect freedom of science, and not affect scientific research and development or testing of models before they are put into market or use. Only military, national security and defence are completely exempt from having to conform to the AI Act regulation (Definition 24 \cite{EU}). 

The aforementioned aspects of the EU AI Act add a lot of burden to the developer and/or provider as the system cannot make the jump from research into production without conforming to the rules. This is of course understandable as it provides much needed guardrails against the risk with AI such as rampant usage of AI in profiling etc. Nonetheless, the regulations also slow down the transition of new research-based systems to real world usage while also adding to the costs of development. In some cases it might be beneficial to have a reduced set of regulations to fulfill in order to release a system for limited specialist usage instead of a general public. 

\section{Conclusions}

We reviewed two AI systems, work disability risk and Alzheimer's disease risk prediction. Both systems seem to have been developed following good ethical standards for both the medical and AI aspects. The EU AI Act classification of both systems is 'high-risk AI' due to categorization of people (cases A and B), and sensitive biomarkers (case B). That high risk classification means many different measures and actions must be taken before the system can be put into production use, and this makes it much more expensive to develop systems. 

The EU AI act makes the self-regulation (such as independent review boards and individual research ethics) from an ethical consideration into a matter of compliance and conformity with regulatory governance, while intended to safeguard and increase accountability, it ends up subornating ethical reflection and professional researchers to legal compliance, undermining responsible research practices and the autonomy of research.   

\paragraph{Acknowledgements}
This work was partly funded by Business Finland under MAISA programme.

%
% ---- Bibliography ----
%
%\bibliographystyle{splncs04}
%\end{document}

\end{document}